\documentclass[12pt]{article}
\usepackage{amsfonts,amsmath}
\usepackage{amssymb}
\usepackage[hypertex] {hyperref}

\makeatletter \@addtoreset{equation}{section} \makeatother

\newcommand{\be}{\begin{equation}}
\newcommand{\ee}{\end{equation}}
\newcommand{\bee}{\begin{eqnarray}}
\newcommand{\beee}{\begin{array}}
\newcommand{\eee}{\end{eqnarray}}
\newcommand{\eeee}{\end{array}}

\newcommand{\un}{{\underline{n}}}

\newcommand{\ga}{\alpha}

\newcommand{\pb}{{\dot{\gb}}}

\newcommand{\gb}{\beta}

\newcommand{\rhs}{{\it r.h.s.} }

\newcommand{\lhss}{{\it l.h.s.} }
\newcommand{\ie}{{\it i.e.,} }
\newcommand{\ls}{\!\!\!\!\!\!}

\newcommand{\go}{\omega}

\newcommand{\by}{{\bar{y}}}

\newcommand{\bp}{{\bar{p}}}

\newcommand{\q}{\,,\qquad}

\newcommand{\dga}{{\dot{\alpha}}}
\newcommand{\dgb}{{\dot{\beta}}}

\newcommand{\nn}{\nonumber}

\newcommand{\half}{\frac{1}{2}}

\newcommand{\p}{\partial}

\newcommand{\f}{\frac}

\newcommand{\dal}{\dot \alpha}

\newcommand{\al}{ \alpha}
\newcommand{\dr}{{\rm d}}

\begin{document}

\begin{flushright}
{\small FIAN/TD/06-22}
\end{flushright}
\vspace{1.7 cm}

\begin{center}
{\large\bf Projectively-Compact Spinor Vertices and Space-Time Spin-Locality
  in  Higher-Spin Theory}

\vspace{1 cm}

{\bf  M.A.~Vasiliev}\\
\vspace{0.5 cm}
{\it
 I.E. Tamm Department of Theoretical Physics, Lebedev Physical Institute,\\
Leninsky prospect 53, 119991, Moscow, Russia}

\end{center}

\vspace{1.2cm}

\vspace {1cm}

{\it $\phantom{MMMMMMMMMMMMMMMMMM}$ To the memory of Mikhail Soloviev}


\vspace{0.4 cm}

\vspace{0.4 cm}

\begin{abstract}
The concepts of compact and projectively-compact spin-local spinor vertices are
introduced.  Vertices of this type are shown to be  space-time spin-local, \ie
their restriction to  any finite subset of fields is space-time local. The known spinor
spin-local cubic vertices with the minimal number of space-time derivatives
are verified to be projectively-compact. This has the important consequence
that spinor spin-locality of the respective quartic
vertices would  imply their space-time spin-locality.
More generally, it is argued that the proper class of solutions of the non-linear
higher-spin equations that leads to the minimally non-local (presumably space-time spin-local)
vertices is represented by the projectively-compact vertices. The related aspects
of the higher-spin holographic correspondence are briefly discussed.

\end{abstract}

\newpage
\tableofcontents

\newpage

\section{Introduction}
\label{intro}

Higher-spin (HS) gauge theory has dramatic history.
It results from a simply looking idea of promoting HS gauge
symmetries associated with free Fronsdal theory \cite{Fronsdal:1978rb}
to the non-linear level.
In old days HS gauge theory was argued not to exist
due to the so-called no-go statements forbidding HS symmetries \cite{cm,Haag,Weinberg:1965rz}
(see  \cite{Bekaert:2010hw} for a review).
Later on
it was shown that the problem  is avoided by going from the flat background to
$AdS$ \cite{Fradkin:1987ks}.

Fundamental properties of HS gauge theories are that they are consistently
formulated in $AdS$ background \cite{Fradkin:1987ks}, contain
higher derivatives in interactions of degrees
increasing with spins \cite{Bengtsson:1983pd,Berends:1984rq,Fradkin:1987ks}, and
involve infinite towers of fields of unlimited
spins \cite{Berends:1984rq,Fradkin:1986ka}.
Altogether, these properties imply that HS gauge theory is not a usual local theory,
exhibiting certain degree of non-locality. The level of non-locality of HS gauge theory is still debatable in the literature within various formalisms. In this paper we elaborate a
tool reducing the analysis of spin-locality of the HS theory in space-time to the much simpler
one  in the auxiliary spinor space.

In eighties,  the role of $AdS$ background was not appreciated
so that its relevance was considered just as a peculiarity of HS theory.
However, after $AdS/CFT$ came into the game \cite{Maldacena:1997re, Gubser:1998bc,Witten:1998qj}
Klebanov and Polyakov
 proposed a remarkable conjecture that HS theory is holographically dual
 to a $3d$ vector
boundary sigma model \cite{KP} see also
\cite{Konstein:2000bi,Sundborg:2000wp,WJ,Mikhailov:2002bp,Sezgin:2002rt}
for related precursor work.
Further generalizations were worked out to supersymmetric \cite{LP,Sezgin:2003pt}
and Chern-Simons extensions    \cite{Aharony:2011jz,Giombi:2011kc}.

Though Klebanov-Polyakov conjecture
  obeyed all kinematic constraints of the linearized holography
expressed by the Flato-Fronsdal theorem \cite{FF} establishing the relation between
currents (tensor products) build from free $3d$ conformal  fields
and free massless fields in $AdS_4$, attempts to
reconstruct HS interactions in the bulk led the authors of \cite{Sleight:2017pcz,Ponomarev:2017qab}
(see also \cite{Bekaert:2015tva}) to the conclusion
that HS gauge theory must be essentially non-local beyond the leading order.
(See, however, \cite{David:2020ptn}).
This point is wide spread these days even though, as discussed below, it has no
solid grounds whatsoever\footnote{A couple of years ago one very good physicist
and friend of mine told me at some conference: "You know, someone whose name I
even do not remember, told me that HS gauge theory has been shown to be
very poorly defined as the interacting theory being too non-local".}.

Let us now summarise what is really known on the status of the HS gauge theory.
There are two main ways for its study.
The most popular and seemingly simpler one is to analyse the properties of HS
gauge theory based on the holographic duality principle
\be
\label{KP}
KP\quad conjecture \overbrace\Longrightarrow^{Holography} HS \quad gauge\quad
 theory\,.
\ee
This was used for obtaining both  positive results in \cite{Giombi:2009wh,Giombi:2012ms}
 and negative ones in
 \cite{Sleight:2017pcz}.

Alternatively, HS gauge theory can be studied directly in the bulk.
This is hard at higher orders
because of the presence of infinite towers of dynamical fields and the absence of
the low-energy expansion parameter, that demands all higher derivatives be accounted
on the same footing in the $AdS$ background. Apart from standard Noether procedure
typically efficient at the lower-orders \cite{Berends:1984rq} (for incomplete list of references
see also
\cite{Bekaert:2005jf,Bengtsson:2006pw,Manvelyan:2010jr,Sagnotti:2010at,Fotopoulos:2010ay,
Vasilev:2011xf,Joung:2012rv,Buchbinder:2012xa,
Francia:2016weg,Buchbinder:2017nuc}) but not beyond, there are two
main tools to simplify the problem.

One is the light-cone formalism of \cite{Bengtsson:1983pd}
 further developed in the series of papers of A.~Bengtsson (see  \cite{Bengtsson:2012jm}
 and references therein) and Metsaev
\cite{Metsaev:1991mt,Metsaev:1993,Metsaev:1999ui,Metsaev:2006,Metsaev:2007rn,Metsaev:2022yvb},
who in particular was able to find some quartic HS vertices in Minkowski space
\cite{Metsaev:1991mt}  that correspond to the
self-dual HS theory, and others (see e.g. \cite{Ponomarev:2016lrm}).

An efficient covariant approach was suggested in
\cite{Vasiliev:1992av} where a system of equations was formulated allowing to
reconstruct on-shell HS vertices order by order. This way, using shifted homotopy approach,
in
\cite{Vasiliev:2016xui,Gelfond:2017wrh,Gelfond:2018vmi, Didenko:2018fgx,Didenko:2019xzz,Gelfond:2019tac,Didenko:2020bxd,
Gelfond:2021two}\footnote{Note that  the results of \cite{DeFilippi:2019jqq,DeFilippi:2021xon,Iazeolla:2022dal}  suggested that the
shifted homotopy approach can be related to the star-product deformation as was
explicitly demonstrated
in \cite{Didenko:2020bxd}.}
some higher-order vertices
in HS theory were reconstructed  that all have been shown to be spin-local
which roughly speaking means their locality for any finite subset
of fields in the system. (For more detail see \cite{Gelfond:2019tac} and below.) However, to compare
conclusions of the bulk analysis of the equations of \cite{Vasiliev:1992av}
 with the holographic reconstruction conclusion of  \cite{Sleight:2017pcz}
 one has to find full $\phi^4$
vertex with $\phi$ being a scalar field component contained in the zero-form sector of HS fields
$C$ in HS theory. This problem has been only partially solved so far.

The scheme of \cite{Vasiliev:1992av} is complete in the sense that
it represents any solution to the problem including all possible field
redefinitions. This has both advantages and disadvantages. The disadvantage
is that equations of \cite{Vasiliev:1992av} do not directly lead to a local
or minimally non-local form of the equations. In other words, to proceed
one has to work out appropriate additional conditions  that single out
a proper form of HS field equations. This is somewhat analogous
to the Schroedinger equation in quantum mechanics: most of its
solutions  have no relation to physics.
One has to, first, impose an additional condition that the wave-function must
belong to $L_2$ and, second, to find  appropriate solutions.
Analogously, in the analysis of HS gauge theory based on the equations of
\cite{Vasiliev:1992av} a list of conditions has been identified in
 \cite{Gelfond:2018vmi} that reduce the degree of non-locality of the
resulting vertices. As a result, it was
shown that all lowest-order
vertices are properly reproduced by the equations of \cite{Vasiliev:1992av}.
These results were  then extended \cite{Gelfond:2019tac} to a class of
non-linear vertices of the types discussed in \cite{Sleight:2017pcz}. Though
so far not all of the HS vertices
to be compared with \cite{Sleight:2017pcz} have been obtained from
\cite{Vasiliev:1992av} the already obtained results   indicate that the analysis of
\cite{Vasiliev:1992av} along the lines of \cite{Gelfond:2019tac} leads to the
conclusions essentially different from those predicted in
\cite{Sleight:2017pcz}.

It should be noted that the  results of this paper
further restrict a class of solutions of the HS equations of \cite{Vasiliev:1992av}
 leading to space-time spin-local HS dynamics. We conjecture that the projectively-compact
 spin-local vertices identified below just form the appropriate class of solutions of the
 non-linear HS equations of \cite{Vasiliev:1992av} analogous to $L_2$ in QM.

It is important that HS holography is weak-weak \cite{KP,Giombi:2012ms}, \ie it can  be
tested on the both sides in the weak coupling regime. In this
situation disagreement of the bulk and boundary analysis implies that
something goes  wrong in the holographic correspondence.
Since HS gauge theory is definitely one of the fundamental theories of nature,
this has to be taken seriously because its analysis can affect the standard paradigm of
the holographic correspondence in presence of HS gauge fields.

In this paper we  focus on the interplay between the notions of spin-locality
in space-time and in the twistor-like spinor space. (Both  are
explained in the main text.) In practice, it is easier to analyse spinor spin-locality
controlled by a number of theorems of \cite{Gelfond:2018vmi,Gelfond:2019tac}. However, beyond the lowest order,
spinor spin-locality does not necessarily imply spin-locality in space-time that makes it difficult
to compare conclusions
of the analysis of the HS theory in the spinor space against the space-time
locality analysis. In this paper  a simple sufficient criterion is presented guaranteeing that spinor spin-locality
implies space-time spin-locality. It is also shown that this criterion is fulfilled by the local HS vertices obtained in \cite{Vasiliev:2016xui,Gelfond:2017wrh} hence implying that these vertices
and their higher-order extensions are spin-local in the usual space-time sense as well.

Yet hypothetical spin-locality of the bulk HS theory suggests the need
of a modification of the Klebanov-Polyakov conjecture. This can be achieved by replacing
the boundary sigma-model
  respecting  global HS symmetries by its gauged version of
the sigma-model interacting with the conformal HS gauge theory at the $3d$
boundary. This idea was put forward in \cite{Vasiliev:2012vf}
 being motivated by the direct
holographic correspondence  resulting from the so-called unfolded formulation originally introduced in
 \cite{Vasiliev:1988sa}. The advantage from the
HS theory perspective was the manifest control of HS gauge symmetries on the both sides
of the holographic correspondence. We believe that this key
feature of the HS gauge theory has to be respected by any scheme.

The proposal of \cite{Vasiliev:2012vf} is  non-standard in several
respects. As usual, this type of holographic correspondence relates $AdS$
bulk symmetry with the boundary conformal symmetry. However, the boundary
conformal theory is not a standard  CFT  since
conformal gravity and its HS extensions possess no gauge invariant stress tensor.
As a result correlators in this theory should not necessarily respect conformal
bootstrap which assumption was instrumental for the arguments of
\cite{Sleight:2017pcz}. The prise is however that the boundary theory is much
more involved than usual CFT. ($3d$ theories of this class were considered in
 \cite{Nilsson:2015pua,Linander:2016brv}.) Another distinction
   was that the proposal of \cite{Vasiliev:2012vf} seemingly
 leaves unclear how to identify the sources
 for conformal fields once the HS gauge fields are engaged to be dynamical
 (rather than sources as in the conventional approach).
This point will be briefly discussed in Section \ref{bbc}.

The rest of the paper is organized as follows. In Section \ref{locc} we discuss
the peculiarities of the notion of locality in the models with infinite towers of
fields like in HS theory. The notion of space-time spin-locality is introduced
and the role of {\it compact} spin-local field redefinitions is emphasized.
In Section \ref{free fields} the unfolded formulation of free massless fields in $AdS_4$
 is recalled. The notion of spinor spin-locality
is introduced in Section \ref{spin} while the class of {\it projectively-compact} spin-local vertices
is introduced in Section \ref{cond} where it is shown that
such vertices imply space-time spin-locality. The related aspects
of the HS holographic correspondence are briefly discussed in Section \ref{bbc}. Conclusions
are summarized in Section \ref{con}.

\section{Locality, spin-locality and non-locality}
\label{locc}

Let us now discuss peculiarities of  the notions of locality and non-locality
in field theories like HS gauge theory.
Recall that HS gauge theory contains   interaction
 vertices with higher derivatives both in Minkowski \cite{Bengtsson:1983pd, Berends:1984rq}
 and  $AdS$ \cite{Fradkin:1987ks}  backgrounds (for a broader class of HS models
 see also  \cite{Metsaev:2007rn}). The same time, HS gauge theories in $d\geq 4$
 were known to describe infinite towers of fields of different spins
 \cite{Berends:1984rq} because HS symmetries \cite{Fradkin:1986ka}
 are infinite-dimensional. Since the order of maximal  derivatives $ord(V)$
  in a HS vertex $V(s_1,s_2,s_3)$
 for three fields  with spins $s_1,s_2,s_3$  increases with
 involved spins the theory contains an infinite number of derivatives
 once all spins are involved, thus being non-local in the standard sense.
 However, in such theories  there are more
options to be  distinguished.

\subsection{Interactions}

 Let some system describe  fields $\phi_s^A$ characterized by
 quantum numbers called spin $s$ and some Lorentz indices $A$ like
 tensor, spinor, etc. Consider field equations of the  form
$$
E_{A_0,s_0}(\p, \phi)=0\q E_{A_0,s_0}(\p ,\phi) =\sum_{k=0,l=1}^\infty
 a^{n_1\ldots n_k}_{A_0\,A_1\ldots A_l} (s_0,s_1,s_2,\ldots\,, s_l)
 \p_{n_1}\ldots \p_{n_k}\phi_{s_1}^{A_1}\ldots\phi_{s_l}^{A_l}\,.
$$
Here derivatives $\p_n:= \frac{\p}{\p x^n}$ may hit any of the fields $\phi_{s_k}^{A_k}$ with
$s_0$ being the spin of the field on which the linearized  equation is imposed.
{Locality of the equations
 will be treated  perturbatively, \ie
independently at every order $l$.
In {usual perturbatively local} field theory the total number of
derivatives is limited at any order $l$ by some $k_{max}(l)$:}
\be
\label{loc}
a^{n_1\ldots n_k}_{A_0\ldots A_l} (s_0,s_1,s_2,\ldots s_l) =0 \quad \mbox{at}\quad
k>k_{max}(l)\,.
\ee

This condition can be relaxed to {\it space-time spin-locality} condition
\be
\label{sloc}
a^{n_1\ldots n_k}_{A_0\ldots A_l} (s_0,s_1,s_2,\ldots s_l) =0 \quad \mbox{at}\quad
k>k_{max}(s_0,s_1,s_2,\ldots s_l)
\ee
with some $k_{max}(s_0,s_1,s_2,\ldots s_l)$ depending on the spins in the vertex.
In the theories with the finite number of fields where $s$ takes at most a
finite number of values, the conditions (\ref{loc}) and (\ref{sloc}) are
equivalent. However in the HS-like models, where spin $s$
can take an infinite number of values, the locality and spin-locality
restrictions differ. Both  types of theories
have to be distinguished from the genuinely non-local ones in which
there exists such a subset of spins $s_0,s_1,s_2,\ldots s_l$ that (\ref{sloc})
is not true, \ie no finite $k_{max}(s_0,s_1,s_2,\ldots s_l)$  exists at all.

The relaxation of the class of local field theories with the finite number of
fields to the spin-local class is the simplest appropriate  for the
models involving infinite towers of fields. However, it makes sense to further
specify the concept of spin-local vertices as follows.

We call a spin-local vertex {\it compact} if
$ a^{n_1\ldots n_k}_{A_0\,A_1\ldots A_l} (s_0,s_1,s_2,\ldots\,,s_k +t_k\,,\ldots, s_l)=0$ at
$t_k> t_k^0$ with some $t_k^0$
 for any $0\leq k\leq l$ and {\it non-compact}
otherwise.  (Note that here the compactness is in the space of spins, not in space-time.)
In HS theory both types of vertices are present. For instance, the
cubic HS vertices $\go*\go$ constructed in \cite{Fradkin:1987ks}, that are built from
the HS gauge connections $\go$, are spin-local-compact
since they are non-zero iff spins $s_0, s_1, s_2$ obey the triangle inequalities
$s_0\leq s_1+s_2$ etc. On the other hand, vertices associated with the conserved
currents built from gauge invariant field strengths like those considered in
\cite{Gelfond:2018vmi} are spin-local non-compact. Indeed, these include in
particular vertices
$ a^{n_1\ldots n_k} (s_0,0,0)$ that describe conserved
currents of any integer spin $s_0$ built from two spin-zero fields.

\subsection{Field redefinitions}
{A class of local field theories with finite sets of fields is invariant under
 perturbatively local field redefinitions}
 \be
 \label{fr}
\phi^B_{s_0}\to \phi^B_{s_0} +\delta \phi^{B}_{s_0}\q \delta \phi^{B}_{s_0} = \sum_{k=0,l=1}^\infty
b^B{}^{n_1\ldots n_k}_{A_1\ldots A_l}(s_0,s_1,\ldots, s_l)
 \p_{n_1}\ldots \p_{n_k}\phi_{s_1}^{A_1}\ldots\phi_{s_l}^{A_l}
\ee
{with at most a finite number of non-zero coefficients $b^B{}^{n_1\ldots n_k}_{A_1\ldots A_l}(s_0,s_1,\ldots, s_l)$
 at any given order.}
It should be stressed that application of a non-local perturbative
field redefinition to a local field theory makes it seemingly non-local.
Other way around, to answer the question if one or another model is
perturbatively local or not it is not enough to check given vertices.
Instead one has to analyse a less trivial problem whether there exists
a non-local field redefinition  transforming  a seemingly non-local model to
the manifestly local form. Once such a field redefinition is found, the model is
shown to be local and should be analysed in this local frame (field variables).
As long as such a field redefinition is unknown it is an open question whether
the model is local or not. It should be stressed that in practice this means
that it may be hard to prove that one or another model is essentially non-local.\footnote{
Note that we only consider field redefinitions expandable into
power series of derivatives. The ill-defined non-localities like $\Box^{-1}$ are
not allowed. Allowing the latter, any theory can be reduced to a free theory
(see, e.g., \cite{Barnich:1993vg})
which option is not interesting.}

If the (spin-)local frame of a model is known, the next question
 is what is the proper class of field redefinitions that leave
the form of vertices perturbatively local or spin-local? In field theories
with a finite number of fields the answer is that these are
perturbatively local field redefinitions involving a finite number of
derivatives at every order.

In the theories with an infinite number of fields the situation is more
subtle. Naively one might think that appropriate field redefinitions in spin-local theories
 are also spin-local. This is not necessarily  true,
however, because the modified vertex
\bee
\delta  E_{A_0,s_0}(\p ,\phi) \ls&&=\sum_{{s_p}=0}^\infty
\sum_{p,k,k'=0,l,l'=1}^\infty\ls
 a^{n_1\ldots n_k}_{A_0\,A_1\ldots A_l} (s_0,s_1,s_2,\ldots,s_p,\ldots\,, s_l)
\\
&& \ls\ls\ls\ls\p_{n_1}\ldots \p_{n_k}\phi_{s_1}^{A_1}\ldots\phi^{A_{p-1}}_{s_{p-1}}
 \phi^{A_{p+1}}_{s_{p+1}}
\ldots  \phi_{s_l}^{A_l}
b^{A_p}{}^{m_1\ldots m_{k'}}_{B_1\ldots B_{l'}}(s_p,s_{l+1},\ldots, s_{l+l'})
 \p_{m_1}\ldots \p_{m_{k'}}\phi_{s_{l+1}}^{B_1}\ldots\phi_{s_{l+l'}}^{B_{l'}}
\,\nn
\eee
may contain an infinite summation over the spin $s_p$ of the redefined field.
If the vertex and field redefinition were spin-local the result of such a
field redefinition can still be non-local and even ill-defined because an
infinite number of terms with the same field pattern and any number of
derivatives may result from the  terms with different
$s_p$.

This difficulty is avoided if  the field redefinition (\ref{fr})
is spin-local-compact in which case the summation over $s_p$ is always
finite and the modified vertex is both well-defined and spin-local. Thus, in the
spin-local theories with  infinite sets of fields a proper counterpart of the local field
redefinitions in usual theories with  finite numbers of fields
is represented by spin-local-compact field redefinitions. One of the important
consequences of this analysis is that non-compact spin-local
field redefinitions at the lower order may produce non-localities at higher
orders. This implies that the choice of field variables in the theories with
infinite towers of fields is a delicate issue from the very first step.
Once a spin-local frame of a theory  is found,  a very
restricted class of spin-local-compact field redefinitions is compatible with
spin-locality.

One of the central problems in  HS gauge theory is to find
its  spin-local frame if exists. In \cite{Gelfond:2021two} (and references
therein) the local frame was found for a number of vertices including some
up to the fifth order. This was achieved in the spinor formalism which we sketch
now.

\section{Free fields unfolded}
\label{free fields}

For simplicity, in this paper we focus on the most elaborated example of $4d$ HS theory.
The idea of our consideration applies to other HS theories as well, including
those in $3d$ \cite{Prokushkin:1998bq} and any $d$ \cite{Lopatin:1987hz,Vasiliev:2003ev}.

\subsection{Free equations in $AdS_4$}
For the vacuum one-form connection $W_0$ of $sp(4)\sim o(3,2)$, that describes $AdS_4$,
\be
\label{AdS}
 W_0 = \half w^{AB} (x) Y_A Y_{B}\q
\dr w^{AB} +w^{AC} C_{CD} w^{DB}=0\,
\ee
with the $sp(4)$ invariant form $C_{AB}$  ($A,B\ldots =1\,,\ldots4$),
the  unfolded system for  free massless fields described by the one-forms
$\go(y,\by; K | x)$ and zero-forms $C(y,\by;K | x)$
  reads as \cite{Vasiliev:1988sa}
\bee \label{tDO}
\quad&& R_1(y,\overline{y};K\mid x) = \frac{i}{4}\left ( \eta \overline{H}^{\dga\pb}
\bar \p_\dga \bar \p_\dgb
{C}(0,\overline{y};K\mid x)k + \bar \eta H^{\ga\gb}\p_{\ga}\p_{\gb}
{C}(y,0;K\mid x)\bar k\right )\,, \\ \label{tDC}
\quad&&
\tilde{D}_0 (C(y,\overline{y};K\mid x)k) =0\,, \label{DC}
\eee
where $w^{AB}=(\go^{\ga\gb},\bar \go^{\dga\dgb}, e^{\ga\dga})$
describes Lorentz connection, $\go^{\ga\gb},\bar \go^{\dga\dgb}$, and vierbein, $e^{\ga\dga}$, with two-component
spinor indices $\ga=1,2$, $\dga=1,2$ (at the convention $A^\ga =\epsilon^{\ga\gb}A_\gb$,
$A_\ga =A^\gb\epsilon_{\gb\ga}$)
\be
\p_\ga :=\f{\p}{\p y^\ga}\q \bar\p_\dga :=\f{\p}{\p \by^\dga}\,,
\ee
involutive Klein elements
$K=(k,\bar k)$ defined to obey
\be\label{hcom}
\{k,y_{\al}\}=0\,,\quad [k,\bar y_{\dal}]=0\,,\quad k^2=1\q
[\bar k,y_{\al}]=0\,,\quad \{\bar k,\bar y_{\dal}\}=0\,,\quad \bar k^2=1\,,
\quad [k\,,\bar k]=0\,,
\ee
$$
H_{\ga\gb}:= e_\ga{}_\dga e_\gb{}^\dga\q
\overline H_{\dga\dgb}:= e_\ga{}_\dga e^\ga{}_\dgb\,,
$$
\be
\label{R1}
R_1 (y,\bar{y};K\mid x) :=D^{ad}_0\omega (y,\bar{y};K\mid x)\qquad
D^{ad}_0  = D^L  -
 e^{\ga\pb}\Big(y_\ga \bar \partial_\pb
+ {\p_\ga}\bar{y}_\pb\Big)\,,
\ee
$$
D^L   = \dr_x  -
\Big(\go^{\ga\gb}y_\ga {\p_\gb} +
\bar{\go}^{\dga\pb}\bar{y}_\dga\bar \p_\dgb \Big)\,,
$$
\be
\label{TD}
\tilde{D}_0  = D^L  + e^{\ga\pb}
\Big(y_\ga \bar{y}_\pb +\p_\ga
\bar \p_\pb\Big)\,.
\ee

The massless fields obey
\be
\go(y,\by;-k,-\bar k\mid x) = \go(y,\by;k,\bar k\mid x)\q
C(y,\by;-k,-\bar k\mid x) = - C(y,\by;k,\bar k\mid x)\,.
\ee

\subsection{Zero-form sector}

The linearized HS equations decompose into independent
subsystems associated with different spins.
We start with the simpler equations (\ref{tDC}) on the gauge invariant zero-forms $C$
$$
C(Y;K|x)=\sum^1_{A=0}\sum_{n,m=0}^\infty\f{1}{2 n!m!}
C^{A\,1-A}_{\ga_1\ldots \ga_n\,,\dga_1\ldots \dga_m}(x)
y^{\ga_1}\ldots y^{\ga_n}
\bar{y}^{\dga_1}\ldots \bar{y}^{\dga_m}k^A \bar k^{1-A}\,.
$$
Spin-$s$  zero-forms are $C^{A\,1-A}_{\ga_1\ldots \ga_n\,,\dga_1\ldots \dga_m}(x)$ with
\be
\label{nms}
n-m=\pm 2s\,.
\ee

Eq.~(\ref{TD}) rewritten in the form
\be\label{xyy}
D^L C^{A\,1-A} = e^{\ga\dgb}{ \frac{\partial^2}{\partial y^\ga
\partial \bar{y}^\pb}} C^{A\,1-A} +
\mbox{ lower-derivative and nonlinear terms}
\ee
{implies that higher-order terms in} $y$ {and} $\bar y$ in the zero-forms $C(y,\by| x)$
(discarding from now on indices $A$)
{describe higher-derivative descendants of the primary components $C(y,0| x)$
and  $C(0,\by| x)$.} Generally, $C_{\ga_1\ldots \ga_n\,,\dga_1\ldots \dga_m}(x)$ contain
$\f{n+m}{2}-\{s\}$ space-time  derivatives of the  spin-$s$ dynamical
fields. In particular, self-dual and anti-self-dual components of the generalized
Weyl tensor $C_{\ga_1\ldots \ga_{2s}}(x)$ and $C_{\dga_1\ldots \dga_{2s}}(x)$
originally introduced by Weinberg \cite{Weinberg:1965rz} (see also \cite{penr})
contain $[s]$ derivatives of the dynamical (Fronsdal) massless field, including
spin-zero and spin-1/2 matter fields. As a result, the presence of zero-forms
$C$ in the HS vertices may induce infinite towers of derivatives and, hence,
non-locality.

Using the frame one-form $e^{\ga\dgb}$, whatever  the \rhs of
equations (\ref{xyy}) is it can be represented in the form
\be
\label{F}
D^L C(y,\bar y|x) = e^{\ga\dga}\big(\p_\ga\bar \p_\dga F^{++}(y,\bar y|x) +y_\ga\bar
\p_\dga F^{-+}(y,\bar y|x)+
\p_\ga \bar y_\dga F^{+-}(y,\bar y|x) + y_\ga \bar y_\dga F^{--}(y,\bar y|x)\big )\,.
\ee
Note that this form of the dynamical equations is not unfolded since
their \rhs may contain via $F^{ab}$  components of the dynamical
fields not necessarily in the form of wedge products. For instance, a one-form
$E(y,\by)=dx^\un E_\un(y,\by)$ is represented by  $E^{\pm\pm}(y,\by)$
with
\be
N_y N_{\by}E^{++}(y,\by) := y^\ga \by^\dga e^\un_{\ga\dga}E_\un (y,\by)\q
(N_y+2) N_{\by}E^{-+}(y,\by) := \p_\ga \by^\dga e^\un{}^\ga{}_{\dga}E_\un (y,\by)\,,
\ee
\be
N_y (N_{\by}+2)E^{+-}(y,\by) := y^\ga \bar \p_\dga e^\un{}_{\ga}{}^{\dga}E_\un (y,\by)\q
(N_y+2) (N_{\by}+2)E^{--}(y,\by) := \p_\ga \bar \p_\dga e^\un{}^\ga{}^{\dga}E_\un (y,\by)
\,,
\ee
\be
N_y:=y^\ga\p_\ga \q N_{\by} := \by^\dga \bar\p_\dga\,,
\ee
where $e^\un_{\ga\dga}$ is the inverse vierbein.
The leading term in (\ref{xyy}), that determines the higher
components in $C(y,\bar y|x)$ via space-time derivatives of the lower
ones with smaller $n+m$, just has the form of $F^{++}(y,\by|x)$ (\ref{F}).
 Clearly, the projector $\Pi^{des}$ to the part $F^{++}$ that contains descendants in (\ref{xyy}) is
\be\label{Pi++}
\Pi^{des} :=N_y^{-1} \bar N_\by^{-1} y^\ga \by^\dga \f{\p}{\p e^{\ga\dga}}\,.
\ee

Suppose now that all non-linear corrections to the \rhs of the field equations
(\ref{xyy}) contain  dependence on $y^\ga$ or $\by^\dga$ in the combinations
$e^{\ga\dga} y_\ga \bar \phi_\dga(y,\by)$ or $e^{\ga\dga} \by_\dga  \phi_\ga(y,\by)$.
In that case the expressions for the components $C_{\ga_1\ldots \ga_n\,,\dga_1\ldots \dga_m}(x)
y^{\ga_1}\ldots y^{\ga_n}
\bar{y}^{\dga_1}\ldots \bar{y}^{\dga_m}$ with higher $n+m$ (descendants) via space-time
derivatives of those with the lower ones will preserve the same form as in the free
theory being insensitive to the non-linear corrections to (\ref{xyy}). In  that case
the non-linear corrections to the unfolded HS equations will contribute to the dynamical
equations (the \rhs of the Fronsdal equations and Bianchi identities associated with
them) not affecting the expressions for descendants in $C(Y|x)$ via space-time
derivatives of the primaries. This simple observation will allow us in Section \ref{cond} to
formulate the equivalence condition for
 the concepts of space-time and spinor spin-locality.

\subsection{One-form sector}
\label{onef}

In the sector of one-forms $\go(Y;K|x)$   spin-$s$ fields are described
by the  degree $s-1$  homogeneous monomials
in $Y$:
$
\go(\mu Y; K|x)= \mu^{2(s-1)} \go(Y;K|x)\,,
$
\ie
the spin-$s$ gauge fields in the generating function
$$
\go(Y;K|x)=\sum_{A=0}^1\sum_{n,m=0}^\infty\f{1}{2 n!m!}
\go^A_{\ga_1\ldots \ga_n\,,\dga_1\ldots \dga_m}(x) (k\bar k)^A
y^{\ga_1}\ldots y^{\ga_n}
\bar{y}^{\dga_1}\ldots \bar{y}^{\dga_m}\,
$$
are $\go^A_{\ga_1\ldots \ga_n\,,\dga_1\ldots \dga_m}(x)$ with $n+m=2(s-1)$. Dynamical HS fields, that
contain Fronsdal fields, are those with $n=m$ for bosons
and $|n-m|=1$ for fermions. Other components contain
\be
\label{nd}
\# (\p_x)=\half (|n-m|-2\{s\})
\ee
derivatives of them.  In other words, dynamical fields  belong to the
bisectrix on the plane $n$, $m$ for bosons or to its nearest neighbours for fermions.
For given $s$ the number of derivatives in the field
$\go_{\ga_1\ldots \ga_n\,,\dga_1\ldots \dga_m}$ (discarding from now on the dependence on
$K$ and $x$) equals to the half a distance
from the nearest line of the dynamical fields on the $n,m$ plane.
This has the important consequence that the spin-$s$ components of $\go(Y)$
contain at most $s-1$ derivatives of the spin-$s$ Fronsdal field.

The interpretation of the components $\go_{\ga_1\ldots \ga_n\,,\dga_1\ldots \dga_m}$
depends on whether $n> m$ or $n < m$. From (\ref{R1}) it follows that at $n>m$ every next
spin-$s$ component with $n>m$ is expressed by (\ref{tDC}) via the space-time derivatives of
the previous one
\be
\label{1der+}
D^L  \go (y,\by) - e^{\ga\pb}
\partial_\ga \bar{y}_\pb\go(y,\by)+\ldots =0\q n\geq m\,,
\ee
where  ellipsis  denotes the lower-derivative terms as well as the \lhss of the
Fronsdal equations or Bianchi identities.
Analogously, at $m\geq n$
\be
\label{1der-}
D^L  \go (y,\by) - e^{\ga\dgb}{y}_\ga
\bar \partial_\dgb\go(y,\by)+\ldots =0\q m\geq n\,.
\ee
Introducing (non-orthogonal) projectors $P_{\pm}$:
\be
\label{P+}
\quad P_+ \go(y,\by) = \go(y,\by)\quad n\geq m \q \quad P_+ \go(y,\by) = 0\quad n< m\,,
\ee
\be
\quad P_- \go(y,\by) = \go(y,\by)\quad m\geq n \q \quad P_- \go(y,\by) = 0\quad m< n\,,
\ee
equations (\ref{1der+}), (\ref{1der-}) can be put into the form
\be
\label{edot}
D^L  \go (y,\by) - e^{\ga\pb}\Big (P_+
\p_\ga\bar{y}_\pb+
 P_- {y}_\ga
\bar \p_\dgb \Big )\go(y,\by)+\ldots =0\,.
\ee
Any one-form $\go(y,\by)$ can be represented in the form
\be
\go(y,\by) =
e^{\ga\dga}\big(\p_\ga\bar \p_\dga \Omega^{++}(y,\bar y) +y_\ga\bar
\p_\dga \Omega^{-+}(y,\bar y)+
\p_\ga \bar y_\dga \Omega^{+-}(y,\bar y) + y_\ga \bar y_\dga \Omega^{--}(y,\bar y)\big )
\ee
with zero-forms $\Omega^{\nu\mu}$, $\nu,\mu= +$ or $-$. It is not hard to see that
\be
\label{ep+}
e^{\ga\pb}
\bar \p_\dgb{y}_\ga \go (y,\by) = \half \Big ((N_\by +2) H^{\ga\gb}\big(y_\ga  \p_\gb \Omega^{-+}(y,\by)+
y_\ga y_\gb \Omega^{--}(y,\by)\big)
-  N_y \bar H^{\dga\dgb} \bar \p_\dga\bar \p_\dgb \Omega^{++}(y,\by)
\Big )\,,
\ee
\be
\label{ep-}
e^{\ga\pb}
\p_\ga\bar{y}_\pb \go (y,\by) = \half \Big ((N_y+2)\bar H^{\dga\dgb}\big (\by_\dga \bar \p_\dgb
\Omega^{+-}(y,\by)+
\by_\dga \by_\dgb  \Omega^{--}(y,\by)\big)
-  N_\by H^{\ga\gb} \p_\ga\p_\gb  \Omega^{++}(y,\by)
\Big )\,.
\ee

Note that the \rhs of equation (\ref{tDO}) also has this form with
$C(y,0;K|x)$ and $C(0,\by;K|x)$ in place of the appropriate components of $\Omega^{++}$.

Suppose that non-linear corrections to the \rhs of the field equations
(\ref{tDO}) do not contribute to the projected terms of (\ref{ep+}) and (\ref{ep-})
$e^{\ga\pb}\Big (P_+\p_\ga\bar{y}_\pb+ P_- {y}_\ga
\bar \p_\dgb \Big )\go(y,\by)$.
Then expressions for the components of $\go(y,\by)$
associated with higher space-time derivatives of the Fronsdal fields
keep the same form as in the free
theory being insensitive to the non-linear corrections to (\ref{tDO}). Hence,
the latter will only contribute to the \rhs of the Fronsdal equations and associated Bianchi
identities not affecting the expressions for higher components of $\go(y,\by|x)$ via space-time
derivatives of the lower ones. This fact  underlies the analysis
of the equivalence of
space-time and spinor spin-locality  in Section \ref{cond}.

\section{Spinor spin-locality}
\label{spin}

Unfolded HS equations have the following  form originally proposed in
\cite{Vasiliev:1988sa}
\begin{equation}\label{HSsketch1}
\dr_x \omega+\go\ast \go=\Upsilon(\go,\go,C)+\Upsilon(\go,\go,C,C)+\ldots,
\end{equation}
\begin{equation}\label{HSsketch2}
\dr_x C+\omega \ast C-C\ast \omega=\Upsilon(\go,C,C)+\Upsilon(\go,C,C,C)+\ldots\,.
\end{equation}
 As in \cite{Vasiliev:1988sa}, the perturbative
expansion is in powers of the zero-forms $C$ with one-forms $\go$ treated as having order zero.
Only wedge products of differential forms are used with the wedge symbol implicit.

Note that from the above results it is not hard to check that
the vertices (\ref{HSsketch1}) and (\ref{HSsketch2}) that are
at most linear in the zero-forms $C$ are spin-local and, moreover,
spin-local-compact. On the other hand, the vertices bilinear in $C$
are {\it a priori} not spin-local and may contain infinite towers
of derivatives for any given set of spins. Nevertheless, in the papers
\cite{Gelfond:2018vmi, Gelfond:2019tac, Gelfond:2021two} a field frame
was found in which these vertices are manifestly spin-local though non-compact.

As explained in Section \ref{free fields},  equations
(\ref{xyy}) tell us that, in the lowest order, $\f{\p}{\p x}$ is equivalent to
$\f{\p^2}{\p y \p \bar y}$. However, at higher orders  relation
(\ref{xyy}) gets corrected due to the contribution of the non-linear
vertices to the unfolded HS equations at lower orders.

The customary field-theoretic setup is that of space-time
derivatives. From the perspective of non-linear HS  equations the analysis
of locality in terms of spinor variables $y$ and $\bar y$ is most
natural, being technically far simpler and allowing simple criteria of spinor spin-locality elaborated in \cite{Gelfond:2018vmi,Gelfond:2019tac}. Since the two approaches are different we will call them differently
using the name {\it space-time spin-locality} for the former and {\it spinor spin-locality}
for the latter.\footnote{Note that  earlier
the term spin-locality was sometimes used for slightly different concepts \cite{Gelfond:2018vmi}.}

Let us now explain how the spinor spin-locality works \cite{Gelfond:2018vmi}.
The vertices {can be put into the form}
\be\label{vert}
\Upsilon(C,C,\ldots)=
F(y, t^i , p^{l}, \bar y; \bar t^j \bar p^{k})\go(Y_1)\ldots \go(Y_k) C(Y_{k+1})\ldots C(Y_n)\Big |_{Y_i=0}\,,
\ee
where
$$
t^i_\ga := \f{\p}{\p y^\ga_i}  \q \bar t^i_\dga := \f{\p}{\p \bar y^\dga_i}
$$
act on the argument of the $i^{th}$ factor of $\go$ and
$$
p^i_\ga := \f{\p}{\p y^\ga_i}  \q \bar p^i_\dga := \f{\p}{\p \bar y^\dga_i}
$$
act on the argument of the $i^{th}$ factor of $C$.
The function $F(y, t^i , p^{l}, \bar y; \bar t^j \bar p^{k})$
depends on various Lorentz-covariant contractions of $y_\ga, p^i_\ga$ and $t^i_\ga$ and their conjugates like
\be\label{svertkiTP}
P^{ij}:=p^i_\ga p^{j\ga}\q
\bar P^{ij}:= \bar p^i_\dga \bar p^{j\dga}\q
T^{i}:=y^\ga t_\ga^i\q \bar T^i:= \bar y^\dga \bar t^i_\dga\,.
\ee
Generally, as a consequence of non-locality of the (Moyal) star product underlying the full
nonlinear system \cite{Vasiliev:1992av}, once no special care of the
perturbative scheme is taken, the functions
$
F(P^{ij}, \bar P^{kl})\,
$
are non-polynomial. Since an infinite number of derivatives $p^i_\ga$ implies
by (\ref{xyy}) an infinite number
of space-time derivatives, such a general vertex is non-local.

Note that there was a number of studies of the star-product induced non-locality in the
literature (see e.g. \cite{3,Soloviev:2014agi,Soloviev:2019keq,Bekaert:2021sfc}). However, the
HS setup of \cite{Vasiliev:1992av}  provides special
tools to control locality in HS theories as explained in \cite{Vasiliev:2015wma,Gelfond:2018vmi,Gelfond:2019tac} and in this paper.

 Since a spin-$s$ one-form connection $\go$  contains
at most a finite number of derivatives dominated by $s$,
from the spinor spin-locality perspective the potential non-local contribution
to the vertices is due to the zero-forms $C$.
If, however, there exists such a perturbative scheme that
 $
F(P^{ij}, \bar P^{kl})\,
$
is polynomial
{in either} $P^{ij}$ {or} $\bar P^{ij}$ for any pair of  $ i,j$,
the vertex is called spinor spin-local. In the lowest order this implies space-time spin-locality.
Indeed, since the
{projection on the fixed spins relates degrees in} $P^{ij}$ {and}
 $ \bar P^{ij}$ {to each other} via (\ref{nms}), if the function was polynomial in
 $ P^{ij}$, its projection to  fixed spins is simultaneously polynomial
 in $ \bar P^{ij}$ and vice versa because the degrees in $y$ and $\by$  differ
  by $2s$.

Remarkably the perturbative scheme elaborated in
\cite{Didenko:2018fgx,Didenko:2019xzz,Gelfond:2019tac,Didenko:2020bxd,Gelfond:2021two} reproduces spinor spin-local vertices of this type
not only in the $C^2$ sector but also in the holomorphic $\Upsilon^{\eta\eta}
(\go,C,C,C)$
and antiholomorphic $\Upsilon^{\bar\eta\bar\eta}
(\go,C,C,C)$
parts of the vertex $\Upsilon(\go,C,C,C)$ proportional to $\eta^2$ and
$\bar\eta^2$, respectively where $\eta$ is the coupling constant in the HS equations of
\cite{Vasiliev:1992av} (for review see \cite{Vasiliev:1999ba}).\footnote{Recently,
 a closely related  attempt of the analysis of holomorphic vertices pretending to be unrelated
 to equations of \cite{Vasiliev:1992av} and the approach of
 \cite{Didenko:2020bxd} (though using the star-product of \cite{Didenko:2020bxd})
 was suggested
in \cite{Sharapov:2022awp} where however neither its formal consistency nor potential divergencies
were considered.}
Once these results are extended to the mixed vertex
$\Upsilon^{\eta\bar\eta}
(\go,C,C,C)$ this proves spin-locality of the $C^3$ ($C^4$ at the action level)
that was argued to be non-local in \cite{Sleight:2017pcz}. At the moment this is still
work in progress.

\section{Space-time : spinor spin-locality equivalence\\ for projectively-compact vertices}
\label{cond}

Though spinor and space-time spin-locality are equivalent in the lowest order this is
not necessarily true at higher orders. The reason is that non-linear contributions to
the field equations (\ref{xyy}) may affect the expressions for higher components of the
zero-forms $C$ via space-time derivatives of the lower ones adding some higher-derivative
non-linear terms to the former. In that case, spinor spin-locality will not
imply the space-time one.  There exist, however, a special class of {\it projectively-compact} vertices for which
 this does not happen and spinor spin-locality implies the space-time one. Remarkably, as discussed below,
such vertices  have been already
identified in  HS theory as carrying the minimal number of derivatives.

\subsection{Spin-local-compact vertices}
If an order-$n$ vertex is spinor spin-local-compact it will only
produce space-time spin-local vertices. This is simply because
for a given set of spins at most a finite
number of higher-derivative terms will contribute to the next-order space-time vertex
in full analogy with the
spin-local-compact field redefinitions (\ref{fr}). For instance, the redefinition
of the descendant HS fields due to the presence of the
spin-local-compact vertices $\go*\go$ and
$\go\go C$ do not spoil space-time spin-locality at the next order.
Though the situation with the non-compact spinor spin-local vertices $V$ containing several factors
of the zero-forms $C$ is more subtle it can be controlled in a similar way in case
its projection $\Pi^{des} V$ to the sector determining the descendant fields in terms of the
derivatives of the primary ones (cf. (\ref{xyy}), (\ref{Pi++})) is spin-local-compact. We will call
such vertices {\it projectively-compact spin-local}.
In particular, this is true if $\Pi^{des} V=0$.

\subsection{Projectively-compact spin-local vertices in $\dr_x C$}
Let us start with the simpler case of  equation (\ref{HSsketch2}) on the zero-form $C$.
Suppose that the vertex obeying the  spinor spin-locality conditions of Section \ref{spin}
has the form (\ref{vert}) with
 \be
 \label{Fpr}
F(y, t^i , p^{l}, \bar y; \bar t^j \bar p^{k}) =
\tilde T  F'(P^{ij}, \bar P^{kl},\ldots)
\ee
with
$\tilde T= T$ or $\bar T$ (\ref{svertkiTP}). Then non-linear corrections resulting from this vertex
do not affect the space-time spin-locality   at the next order.
To see this one has to take into account that in the lowest order the one-form $\omega$
contributes via its zero-order background part associated with the vierbein. The presence of
the factor of $\tilde T$ implies that each vierbein comes
 via $e_{\ga \dga} y^\ga$ or $e_{\ga \dga} \bar y^\dga$. However, as explained in Section
\ref{free fields}, such terms do not contribute to $F^{++}$ in (\ref{F}) that
determines the expressions for higher components in spinor variables in $C(Y|x)$ via
space-time derivatives of the lower ones. So, in this case, spinor spin-locality
with $\Pi^{des} $ (\ref{Pi++})
is equivalent to the space-time spin-locality at least up to the higher-order contributions.

The vertex  found in \cite{Vasiliev:2016xui}
\be
\label{ex}
\Upsilon = \Upsilon_{\eta }(e,C)
 + \Upsilon_{\bar \eta}  (e,C)
\ee
has the following projectively-compact spin-local form presented in \cite{Gelfond:2017wrh}
 \be\label{hh'}
 \Upsilon_{\eta } (e,C)=
 \frac{1}{2}\eta
  \exp{(i  \bar P^{1,2} )}
\int_0^1 d\tau e( y, (1-\tau) \bp_1-\tau\bp_2  )
C(\tau y , \bar y ;K) C(-(1-\tau) y , \bar y   ;K) *k\,,
\ee
\be\label{barhh'}
 \Upsilon_{\bar \eta}  (e,C)=
  \frac{1}{2}{\bar \eta}   \exp i  ( P^{1,2})
\int_0^1 d\tau e((1-\tau)  p_1- \tau   p_2{}  ,\by)
C(   y   , \tau \bar y;K)C( y ,  -(1-\tau) \bar y;K ) *\bar k\,,
\ee
where $e^{\ga\dga}$  is the $AdS_4$ vierbein, $\eta$ is a free complex parameter of the HS theory, and
\be
e(a,\bar a): = e^{\ga\dga}a_\ga \bar a_\dga\,.
\ee

It is spinor spin-local since either $P^{12}$ or $\bar P^{1,2}$ enter non-polynomially
 but not both. Since the frame field $e^{\ga\dgb}$ is contracted either with
$y$ or to $\bar y$ being of the form (\ref{Fpr}) it is also projectively-compact
not affecting
the interplay between spinor and space-time spin-locality at the next order.\footnote{Note that,
since the cubic spin-zero vertex is zero in the HS theory \cite{Sezgin:2003pt,Gelfond:2010pm},
the relation between descendants and primary fields is not affected by the interactions
 in this sector. This observation implied \cite{misuna} equivalence of the space-time and
spinor spin-locality for the quartic spin-zero vertex analysed in \cite{Sleight:2017pcz}.}
This implies in particular  that the spinor spin-local $C^3$
vertices of \cite{Didenko:2019xzz,Didenko:2020bxd,Gelfond:2021two} are space-time
spin-local  provided that one starts with
$C^2$ vertices (\ref{hh'}), (\ref{barhh'}).

The following comment is now in order. A special property of the vertex
(\ref{Fpr}) is  that it has a lowest number of space-time derivatives compared to
 other vertices containing  factors like $t^i_\ga p^j{}^\ga$ in place of $t^i_\ga y{}^\ga$. This is because the
replacement of $y$ by $p$ increases the degree of the respective component of the zero-form
$C$ which by virtue of (\ref{xyy}) implies increase of the number of space-time derivatives carried by
 $C$. The lesson is that, to reach equivalence between
space-time and spinor spin-locality, one has to choose the vertices with the minimal number of
derivatives among various local ones. Note that the vertex of \cite{Vasiliev:2016xui}
was successfully tested in \cite{Didenko:2017lsn,Sezgin:2017jgm} to show that it
properly reproduces the  holographic predictions.

\subsection{Projectively-compact spin-local vertices in $\dr_x \go$}

As explained in the end of Section \ref{onef}, the terms in the vertex that have the form
\be
\label{ep+'}
 P_+  \Big ((\bar N+2) H^{\ga\gb}(y_\ga  \p_\gb \Omega^{-+}(y,\by)+
y_\ga y_\gb \Omega^{--}(y,\by))
-  N \bar H^{\dga\dgb} \bar \p_\dga\bar \p_\dgb \Omega^{++}(y,\by)
\Big )\,
\ee
or
\be
\label{ep-'}
P_-  \Big ((N+2)\bar H^{\dga\dgb}(\by_\dga \bar \p_\dgb \Omega^{+-}(y,\by)+
\by_\dga \by_\dgb \Omega^{--}(y,\by))
-  \bar N H^{\ga\gb} \p_\ga\p_\gb  \Omega^{++}(y,\by)
\Big )\,
\ee
with $P_{\pm}$ (\ref{P+})
do not affect the expressions for the one-form descendant fields via derivatives of
the primary ones hence being projectively-compact.
In that case, spinor spin-locality of the next-order vertex implies its space-time spin-locality.

Remarkably, the cubic vertices found in \cite{Gelfond:2017wrh} do indeed have such a form. Moreover,
they only contain the $y,\bar y$-independent terms with nonzero $\Omega^{++}$.
(For explicit expressions that indeed contain the projectors $P^{\pm}$ in the form of appropriate Cauchy
integral see  Eq.~(3.7) of \cite{Gelfond:2017wrh}.)
In fact, this again implies that the respective vertices carry the minimal number of derivatives.

\section{Bulk to boundary correspondence}
\label{bbc}

The idea of the  bulk-to-boundary correspondence suggested by the
unfolded dynamics analysis of \cite{Vasiliev:2012vf} consists of the
observation that, upon an appropriate change of the reality conditions
on spinor variables,  the unfolded equations (\ref{tDC}) make sense independently
of how many space-time coordinates are involved. In the case of $AdS_4$
with local coordinates $x^{\ga\dgb}$ and $AdS_4$ connection $W_0$
these equations describe massless fields in $AdS_4$. The same equations
with the $3d$ local coordinates $\mathbf{x}^{\ga\gb}= \mathbf{x}^{\gb\ga}$  and $3d$ flat
$O(3,2)$ connection describe $3d$ conformal conserved currents. This is a
manifestation of the observation of \cite{Gelfond:2010pm} that the rank-two unfolded
equations with the doubled spinor variables describe conformal conserved
currents (which provides a field-theoretic realization of the Flato-Fronsdal
theorem). The HS connections $\go(Y;K|x^{\ga\dgb})$ describe HS gauge fields
in $AdS_4$ or $3d$ conformal  HS gauge fields $\go(Y;K|\mathbf{x}^{\ga\gb})$
originally considered in \cite{Fradkin:1989xt} (in the absence of $K$, however).

Analogously, the $4d$ system of non-linear HS equations of \cite{Vasiliev:1992av}
can be reinterpreted as a $3d$ non-linear system that describes interactions
of $3d$ conformal currents  with $3d$ conformal HS gauge fields of the
Fradkin-Tseytlin type \cite{Fradkin:1985am,Segal:2002gd}, which however carry no local degrees of freedom
analogously to  $3d$ gravity \cite{d3grw} and supergravity \cite{d3gr}. That allowed us
to conjecture in \cite{Vasiliev:2012vf} that the HS gauge theory in $AdS_4$
is holographically dual to the $3d$ conformal HS theory that describes
interactions of $3d$ conformal currents with the 3d conformal HS gauge fields.
It was shown in \cite{Vasiliev:2012vf} how this picture is reproduced in the
boundary limit $z\to 0$ with $z$ being the Poincar\'e coordinate.

This conjecture can be considered as a HS gauged version of the
 original Klebanov-Polyakov conjecture \cite{KP}.
 In particular, the boundary conformal HS currents
of \cite{Vasiliev:2012vf} can be realized as bilinears
$
J(y_1,y_2|\mathbf{x}) = \sum_{i=1}^N \phi_i(y_1|\mathbf{x}) \phi_i(y_2|\mathbf{x})
$
that obey the same unfolded equations \cite{Gelfond:2003vh}
 hence describing the same
representation of the conformal group. The $N\to\infty$ limit guarantees that
the currents are not subjected to additional constraints.

The important point is that the HS gauge fields $\go$ behave at the boundary as the
shadow fields for the currents represented by the zero-forms\footnote{This is true for spins
$s\geq 1$. For spins $s=0$ or $1/2$ the prescription of \cite{KP} is unchanged  with the
sources and currents distinguished by the boundary behaviour of the respective components
of the zero-forms $C$.} $C$. Therefore, the holographic
prescription is conjectured to be modified with the correlators reproduced by
the variations
\be
\langle J(\mathbf{x}_1)\ldots J(\mathbf{x}_n) \rangle =\f{\delta^n S(\go)}{\delta \go(\mathbf{x_1})\ldots \delta \go(\mathbf{x_n})}
\ee
at the boundary with local coordinates $\mathbf{x}$ of some on-shell gauge
invariant functional $S(\go, C)$ of the type proposed in
\cite{Vasiliev:2015mka}.

\section{Conclusion}
\label{con}
In this paper we introduce a class of {\it projectively-compact spin-local vertices}
establishing equivalence between
space-time and spinor spin-locality (\ie locality of vertices with
any finite set of spins) in HS gauge theory. Spinor
spin-locality of the HS gauge theory is known to be much easier to analyse than
the space-time one. However, generally, the procedure of the derivation of the space-time
vertices from those in the spinor space is not only technically involved but also
may induce higher-derivative corrections beyond the lowest order. Hence, spinor
spin-locality does not automatically imply space-time spin-locality.
In this paper the conditions are found guaranteeing equivalence of the spinor
and space-time spin-locality.

Namely, it is shown that for  so-called projectively-compact spin-local vertices
spinor spin-locality implies the space-time one. It is also checked that
the spin-local vertices of \cite{Vasiliev:2016xui,Gelfond:2017wrh} are
projectively-compact that provides an important step towards verification of the space-time
spin-locality of HS gauge theory via verification of its spinor spin-locality.
We conjecture that the projectively-compact spin-local vertices form a proper functional
class in which HS gauge theory has to be analysed.
Though this is still work in progress  our preliminary results suggest that
HS gauge theory may be  spin-local at the quartic order as well. Once this is shown to be
really true, it would imply space-time spin-locality of the HS gauge theory in
disagreement with the conclusions of \cite{Sleight:2017pcz}  relying on the
Klebanov-Polyakov conjecture \cite{KP}. Most likely, this will demand a modification
of the latter along the lines of \cite{Vasiliev:2012vf} sketched in Section \ref{bbc}.
The most significant new point is that in that case the boundary dual theory is not
just a CFT but rather a conformal HS gauge theory. (Note that as such it does not
possess a gauge invariant stress tensor.) All this can significantly affect
the paradigm of the holographic correspondence replacing gauge - gravity correspondence
by the gravity - conformal gravity one. In particular, this kind of the correspondence spoils the
assumptions of \cite{Maldacena:2011jn}.

For simplicity, in this paper we considered the most elaborated example of $4d$ HS theory.
The idea of our construction applies to other HS theories as well, including
those in $3d$ \cite{Prokushkin:1998bq} and any $d$ \cite{Lopatin:1987hz,Vasiliev:2003ev}.
In the general case the term {\it spinor spin-locality} has to be changed to the {\it fiber spin-locality}
since the fibers may  have non-spinor local coordinates (e.g.,  vectors
in the $d$-dimensional HS theory  of \cite{Vasiliev:2003ev}). Still the rule is that a vertex is
projectively-compact spin-local if its projection to the part of unfolded equations that determines descendant fields
via derivatives of the primary ones is spin-local-compact.

In the end, let us make the following  comment.
Though the analysis of the effect of projectively-compact vertices is perturbative,
it implies space-time spin-locality in higher orders
as well since the corrections to descendants  are projectively-compact spin-local. It
suffices to prove projective spin-locality inductively at any order to prove equivalence of space-time
and fiber spin-locality in all orders.
It is also  worth to mention that both the  notion  of space-time (and, hence, space-time locality)
and spin-locality are intimately related via the underlying symmetry $G$, which is $Sp(4)$
in the $AdS_4$ case. Space-time is  where $G$ is geometrically
realized while spin is associated with the appropriate $G$-modules. Space-time derivatives are
identified with descendants of those modules while spin-locality demands a number of descendants
in the vertex be limited for any finite subset of $G$-modules. Both of the concepts of background geometry
 and spin are perturbative.

\section*{Acknowlegement}
I am most grateful to late Mikhail Soloviev for his interest to the locality
problem in higher-spin gauge theory and many useful  comments. Also I would like to thank Slava Didenko for a useful comment,
 Nikita Misuna for collaboration at the early stage of
the work on the scalar $\phi^4$ vertex and
 Olga Gelfond for many fruitful discussions and  comments
on the manuscript. I wish to thank Ofer Aharony, Theoretical High Energy Physics Group
of Weizmann Institute of Science for the warm hospitality at the final stage of
preparation of this paper. This research was
supported by the Russian Science Foundation grant 18-12-00507.

\end{document}